\documentclass[noshowpacs,showkeys, groupedaddress,twocolumn, floatfix,aps, prb,reprint]{revtex4-1}
\usepackage[english]{babel}
\usepackage{fontenc}
\usepackage{graphicx}
\usepackage{amsmath}
\usepackage{epstopdf}
\usepackage{amssymb}
\usepackage{amsbsy}
\usepackage{amscd}
\usepackage{color}
\usepackage{bbm}
\usepackage{braket}
\usepackage{bm}
\usepackage{siunitx}
\usepackage[normalem]{ulem} 
\usepackage{verbatim}
\usepackage{url}
\usepackage[verbose,hypertexnames=false,bookmarksopenlevel=1,filecolor=blue,
linkcolor=blue,citecolor=blue,urlcolor=blue,pdfstartview=FitH,bookmarksopen,bookmarksnumbered,
colorlinks,plainpages=false,linktocpage]{hyperref}
\DeclareSymbolFont{mathbold}{OML}{cmm}{b}{it}
\DeclareMathSymbol{\bsigma}{\mathord}{mathbold}{27}
\begin{document}
\renewcommand{\figurename}{Fig.}
\title{Magnetoconductance correction in zinc-blende semiconductor nanowires with spin-orbit coupling}
\author{Michael Kammermeier}
\email{michael1.kammermeier@ur.de}
\author{Paul Wenk}
\author{John Schliemann}
\affiliation{Institute for Theoretical
  Physics, University of Regensburg, 93040 Regensburg, Germany}
  \author{Sebastian Heedt}\altaffiliation{Present address: QuTech, Delft University of Technology, 2628 CJ Delft, The Netherlands.}
  \author{Thomas Gerster}\altaffiliation{Present address:
Physikalisch-Technische Bundesanstalt (PTB), Bundesallee 100, 38116 Braunschweig, Germany.}
\author{Thomas Sch\"apers}\affiliation{Peter Gr\"unberg Institute (PGI-9) and JARA-Fundamentals of Future Information Technology, Forschungszentrum J\"ulich, 52425 J\"ulich, Germany}
\date{\today }
\begin{abstract}
  We study the effects of spin-orbit coupling on the magnetoconductivity in diffusive cylindrical semiconductor nanowires.
 Following up on our former study on tubular semiconductor nanowires, we focus in this paper on nanowire systems where no surface accumulation layer is formed but instead the electron wave function extends over the entire cross section.
  We take into account the Dresselhaus spin-orbit coupling resulting from a zinc-blende lattice and the Rashba spin-orbit coupling, which is controlled by a lateral gate electrode.
  The spin relaxation rate due to Dresselhaus spin-orbit coupling is found to depend neither on the spin density component nor on the wire growth direction and is unaffected by the radial boundary.
  In contrast, the Rashba spin relaxation rate is strongly reduced for a wire radius that is smaller than the spin precession length.
    The derived model is fitted to the data of magnetoconductance measurements of a heavily doped back-gated InAs nanowire
and transport parameters are extracted.
At last, we compare our results to previous theoretical and experimental studies and discuss the occurring discrepancies.
\end{abstract}
\maketitle
\allowdisplaybreaks
\section{Introduction}

Despite the intensive research over the past decade, semiconductor nanowires continue to be one of the most active research areas within the nanoscience community. \cite{Yang2010}
The  significant attention is due to the vast scope of technical applications \cite{Greytak2005,Xiang2006,Krogstrup2013,Xing2014,Xing2015}
as well as the capability to conduct fundamental studies, such as the search for Majorana bound states. \cite{Oreg2010,Mourik2012}
Another broad field of interest concerns the utilization for spintronic devices, which exploit the spin degree of freedom of the charge carriers.
The underlying effect, which allows to manipulate the spin, is the spin-orbit coupling (SOC).
To understand the behavior of the spin and control it efficiently, a detailed knowledge about system parameters is essential.
One basic and convenient tool to gather the desired information are weak-field magnetoconductance measurements.
In disordered systems,  the conductivity is either enhanced or reduced due to quantum interference, which is denoted as weak antilocalization (WAL) or weak localization (WL), respectively.
By fitting the experimental data with an appropriate theoretical model, it is possible to extract SOC strengths as well as dephasing, scattering, and, most prominently, spin relaxation rates.\cite{Hansen2005,Dhara2009,XiaoJie2010,
Hernandez2010,Weperen2015,Kammermeier2016,Roulleau2010,Liang2012,Scheruebl2016,Takase2017}
Notably, a crossover from WAL to WL can even indicate spin-preserving symmetries.\cite{Kammermeier2016PRL, Schliemann2016,Kohda2017}

The huge degree of freedom in the device preparation process allows to manipulate many of the nanowire properties  over a wide range.
More precisely, one is able to effectively control  the size, morphology, potential landscape, carrier and impurity concentration, or even crystal structure.\cite{Lauhon2002,Wacaser2006,Fortuna2010,Hernandez2010,Wirth2011,
Bloemers2013,Haas2013,ZhangLu2014,Hollosy2015,Heedt2015,Speckbacher2016} 
Since typically both the SOC and the WAL/WL correction strongly depend on these characteristics, the great diversity makes it difficult to build up a general theoretical description.
In fact, a variety of theoretical models are needed to appropriately characterize the differing nanowires.
The WAL/WL effect in the diffusive regime was analyzed by Kettemann \cite{Kettemann2007a} and Wenk \cite{Wenk2010,Wenk2011} for
planar quantum wires with a zinc-blende lattice.
In our preceding article, Ref.~\onlinecite{Kammermeier2016}, we developed a model for diffusive zinc-blende nanowires where the transport is governed by surface states, which occurs in materials with Fermi level surface pinning \cite{Hernandez2010,Bringer2011,Wirth2011,Heedt2015,Degtyarev2017} or core/shell nanowires. \cite{Bloemers2013}

In this paper, we discuss the situation where the confining potential is flat over the total cross section. 
The motion of the electrons is considered diffusive in three dimensions (3D) and the nanowire radius $R$ much larger than the mean free path $l_e$. 
We take into account the Dresselhaus SOC resulting from the zinc-blende crystal structure and the Rashba SOC, which is controlled by a lateral gate electrode.
Using this, we compute analytically the WAL/WL correction as a function of the nanowire radius.
The spin relaxation rate due to Dresselhaus SOC is found to be independent of the orientation of the spin density, the wire growth direction, and the wire radius.
In contrast, the Rashba spin relaxation rate is strongly reduced for a wire radius that is smaller than the spin precession length $L_{so}$.
If Rashba SOC is present, the long-lived spin densities have a helical structure and, elsewise, are homogeneous in real space.
At last, we fit the derived formulas to experimental data of a heavily doped InAs semiconductor nanowire, which shows a gate-induced crossover from WL to WAL. 
Thereby, we extract spin relaxation and dephasing rates as well as SOC strengths.
To complete our study, we compare our results to the frequently used one-dimensional (1D) magnetoconductance formula of Kurdak \textit{et al.}\cite{Kurdak1992} and discuss the discrepancies in the resulting fits and the gained transport parameters.\cite{Hansen2005,Dhara2009,XiaoJie2010,
Weperen2015,Roulleau2010,Liang2012,Scheruebl2016,Takase2017}

\section{Hamiltonian for bulk electrons}
The Hamiltonian $\mathcal{H}$ which describes bulk electrons in the lowest
conduction band of a zinc-blende type semiconductor with SOC reads as
\begin{align}
\mathcal{H}&=\,\frac{\hbar^2k^2}{2 m}+\mathcal{H}_\text{R}+\mathcal{H}_\text{D}.
\label{nanorodbulk}
\end{align}
The terms
\begin{align}
\mathcal{H}_\text{R}&=\gamma_\text{R}\left[(k_y \mathcal{E}_z - k_z \mathcal{E}_y)\sigma_x+\text{c.p.}\right],\label{rashba}\\
\mathcal{H}_\text{D}&=\gamma_\text{D}\left[k_x(k_y^2-k_z^2)\sigma_x+\text{c.p.}\right], \label{dresselhausZB}
\end{align}
with cyclic permutations (c.p.) of preceding indices, denote the Rashba (R) and Dresselhaus
(D) SOC contributions with the material-specific
parameters $\gamma_i$, the
electric field components $\mathcal{E}_i$, the Pauli matrices $\sigma_i$ and the
effective electron mass $m$.\cite{winklerbook,Zutic2004a,Wu2010} 
In this notation, the underlying basis vectors $\{\mathbf{\hat{x}},\mathbf{\hat{y}},\mathbf{\hat{z}}\}$ point along the crystal axes [100], [010], and [001].
We begin with the assumption that the electrons in the wire experience a nearly homogeneous electric field perpendicular to the wire axis.
In Sec.~\ref{sec:3Dcooperon} we will see that the choice of the wire axis and the perpendicular electric field is arbitrary as the Rashba and Dresselhaus SOC do not mix with each other in the Cooperon and the effect of the Dresselhaus SOC is independent of the crystal direction.
Thus, without loss of generality, we define the wire axis to be oriented along $\mathbf{\hat{z}}$ and the electric field as $\boldsymbol{\mathcal{E}}=\mathcal{E}\mathbf{\hat{y}}$.

\section{Quantum correction to the conductivity}

Within diagrammatic perturbation theory taking into account the quantum interference between self-crossing paths in a disordered conductor gives rise to the first-order
correction to the Drude conductivity $\Delta\sigma$.
The following preconditions on the impurity potential $V_\text{imp}(\mathbf{r})$ are assumed: (i) We consider a standard white-noise
model for the impurity potential, meaning that it vanishes on average
 and is
uncorrelated, i.e., $\left\langle V_\text{imp}(\mathbf{r})\right\rangle=0$ and $\left\langle V_\text{imp}(\mathbf{r})V_\text{imp}(\mathbf{r'})\right\rangle\propto\delta(\mathbf{r}-\mathbf{r'})$, respectively.
(ii) The disorder is weak, i.e., $\hbar/(\epsilon_F \tau_e)\ll 1$, where $\epsilon_F$ is the
Fermi energy and $\tau_e$ is the mean elastic isotropic scattering time.  
Moreover, the motion of the electrons is considered diffusive in all three spatial directions.  
By averaging over all impurities and summing up all maximally crossed ladder
diagrams, we find the quantum correction to the longitudinal static
conductivity~\cite{nagaoka} to first order in $\hbar/(\epsilon_F \tau_e)$ given by the real part of the
Kubo-Greenwood formula
\begin{align}
\Delta\sigma=&\,\frac{2e^2 }{h}\frac{\hbar D_e}{ \mathcal{V}}\Re \text{e}\left(\sum_{\mathbf{Q},s,m_s}\chi_s\braket{s,m_s|\hat{\mathcal{C}}(\mathbf{Q})|s,m_s}\right).
\end{align}
Here, $\mathcal{V}$ is the volume of the nanowire, $D_e$ the
3D diffusion constant, i.e., $D_e=v_F^2 \tau_e/3$, with
the Fermi velocity $v_F$, $\hat{\mathcal{C}}$ the Cooperon propagator, and
$\mathbf{Q}=\mathbf{k}+\mathbf{k'}$ the sum of the wave vector of an electron with spin $\boldsymbol{\sigma}$
and the wave vector of an electron with spin $\boldsymbol{\sigma'}$.
The factor $\chi_s$ is  defined as $\chi_0=1$ and $\chi_1=-1$. 
Furthermore,  the states $\ket{s,m_s}$ represent the  singlet-triplet basis of the system with two electrons, that is, $s\in\{0,1\}$ is the total spin quantum number and $m_s\in\{0,\pm1\}$ the corresponding magnetic quantum number.
We emphasize, that there exists a unitary transformation which associates the triplet basis states with the components of the spin density as shown in Ref.~\onlinecite{Wenk2010} and defined in App.~\ref{app:relation}.
Below, we follow the approach in Refs.~\onlinecite{Kettemann2007a, Wenk2010, Wenk2011,wenkbook,Kammermeier2016} to
compute the quantum correction to the conductivity.
%
\section{3D Cooperon}\label{sec:3Dcooperon}
As SOC constitutes a small perturbation to the kinetic part in the Hamiltonian $\mathcal{H}$ and the main contribution to the Cooperon results from terms near $Q=0$, the Cooperon propagator $\hat{\mathcal{C}}$ can be approximated by 
\begin{equation}
\hat{\mathcal{C}}(\mathbf{Q})=\,\frac{\tau_e}{\hbar}\left(1-\int\frac{d\Omega}{4\pi}\frac{1}{1-i\tau_e\hat{\Sigma}(\mathbf{Q})/\hbar}
\right)^{-1},
\label{cooperon1}
\end{equation}
where $\hat{\Sigma}(\mathbf{Q})=\,\mathcal{H}(\mathbf{Q}-\mathbf{k}_F,\boldsymbol{\sigma})-\mathcal{H}(\mathbf{k}_F,\boldsymbol{\sigma'})$ and the integral is performed over all angles $\Omega$ of the Fermi wave vector $\mathbf{k}_F$.
In 3D the  Fermi contour is nearly spherical and the integral is continuous. 
Since  $\epsilon_F \tau_e/\hbar\gg 1$,
we may further approximate $\hat{\Sigma}(\mathbf{Q})\approx -\mathbf{v}_F(\hbar\mathbf{Q}+2m(\mathbf{\hat{a}}_\text{R}+\mathbf{\hat{a}}_\text{D})\mathbf{S}) $ with the total electron spin vector $\boldsymbol{S}$ in the singlet-triplet basis
as defined in App.~\ref{app:spin}. The matrix $\mathbf{\hat{a}}_\text{R}(\mathbf{\hat{a}}_\text{D})$ contains the contributions due to Rashba (Dresselhaus) SOC, i.e.,
\begin{align}
\mathbf{\hat{a}}_\text{R}&=\,\frac{\alpha_\text{R}}{\hbar}
\begin{pmatrix}
0&0&1\\
0&0&0\\
-1&0&0
\end{pmatrix},
\end{align}
and
\begin{align}
\mathbf{\hat{a}}_\text{D}&=\,\frac{\gamma_\text{D}}{\hbar}
\begin{pmatrix}
k_y^2-k_z^2&0&0\\
0&k_z^2-k_x^2&0\\
0&0&k_x^2-k_y^2
\end{pmatrix},
\end{align}
where $\alpha_\text{R}=\gamma_\text{R} \mathcal{E}$.
For convenience and in analogy to previous publications, Refs.~\onlinecite{Wenk2010,Wenk2011,wenkbook,Kammermeier2016}, we define the Cooperon Hamiltonian $\hat{H}_c=(\hbar D_e \hat{\mathcal{C}})^{-1}$. 
An additional Taylor expansion of the integrand in Eq.~(\ref{cooperon1}) to second order in $(\hbar\mathbf{Q}+2m(\mathbf{\hat{a}}_\text{R}+\mathbf{\hat{a}}_\text{D})\mathbf{S})$, yields 
\begin{align}
\hat{H}_c/Q_{so}^2={}&(\boldsymbol{\mathcal{Q}}+2 e \boldsymbol{\mathcal{A}}_s/\hbar)^2+\lambda_\text{D}\mathbf{S}^2/2,
\label{3DCooperonHamiltonian}
\end{align}
in terms of the dimensionless momenta $\mathcal{Q}_i=Q_i/Q_{so}$ with  $Q_{so}=2m \alpha_\text{R}/\hbar^2=2\pi/L_{so}$ where $L_{so}$ is the Rashba spin precession length. 
Similar to the 2D and quasi-1D cases, \cite{Kettemann2007a,Kammermeier2016} the effect of Rashba SOC becomes manifest in an effective vector potential $\mathbf{A}_s=Q_{so}\boldsymbol{\mathcal{A}}_s$ where $\boldsymbol{\mathcal{A}}_s=\hbar/(2e)(S_z,0,-S_x)^\top$ and therefore couples to the Cooperon momentum.
In contrast, the Dresselhaus SOC leads to a term $\propto\lambda_\text{D}=8\Gamma^2/35$, where $\Gamma=k_F^2\gamma_\text{D}/\alpha_\text{R}$, which does not couple to the wave vector $\boldsymbol{\mathcal{Q}}$ and is diagonal in the triplet sector.
Thus, it gives rise to a spin relaxation rate which is identical for all components of the spin density.
We stress that unlike in tubular wires\cite{Kammermeier2016} the Dresselhaus contribution does not mix with the Rashba contribution and does not depend on the growth direction of the wire due to the averaging over the Fermi contour. 
Hence, the result applies to any zinc-blende nanowire irrespective of the growth direction and for an arbitrarily oriented electric field perpendicular to the wire axis.

\section{Effects of a radial boundary}\label{sec:boundary}
The finite-size geometry of the nanowire requires a boundary condition for the Cooperon.\cite{AltshulerAronov1981,Aleiner2001,Meyer2002,Kettemann2007a}
For an insulating surface and spin-conserving boundary the condition reads as
\begin{align}
\mathbf{\hat{n}}\cdot\left(\boldsymbol{\nabla}+2i e  \mathbf{A}_s/\hbar\right)\hat{\mathcal{C}}\vert_{\mathcal{S}}={}&0,
\label{boundary}
\end{align}
where $\mathbf{\hat{n}}$ denotes the normal vector of the surface $\mathcal{S}$.
This condition accounts for a specular boundary, which is plausible since the nanowires possess only a small degree of surface roughness.\cite{Wang2013a}
Aside from that, in the transverse diffusive regime the ramifications of a diffusive boundary are insignificant.\cite{Khalaf2016} 
The equation above  can be simplified to a Neumann boundary condition, i.e., $\mathbf{\hat{n}}\cdot(\boldsymbol{\nabla}\hat{\mathcal{C}}')\vert_{\mathcal{S}}=0$, by applying a non-Abelian gauge transformation.
Thereby, the Cooperon (and with it the Cooperon Hamiltonian) is transformed as $\hat{\mathcal{C}}\rightarrow\hat{\mathcal{C}}'=U_A\hat{\mathcal{C}}U_A^\dag$ with the unitary transformation operator $U_A=\exp[i 2 e\,(\mathbf{\hat{n}} \cdot\mathbf{A}_s)(\mathbf{\hat{n}}\cdot \mathbf{r})/\hbar ]$.
In this case, the lowest Cooperon mode $\ket{0}$ corresponds to a solution which has a vanishing wave vector perpendicular to the surface, i.e., $\mathbf{\hat{n}}\cdot\mathbf{Q}=0$, and is thus constant in coordinate space along $\mathbf{\hat{n}}$. \cite{Meyer2002}

For a cylindrical nanowire, we identify $\mathbf{\hat{n}}=\boldsymbol{\hat{\rho}}$ and the surface $\mathcal{S}$ is defined by the constraint $\rho=R$ where $R$ is the radius of the wire and we introduced the standard cylindrical coordinates $(\rho,\phi,z)$ with the corresponding basis vectors $\{\boldsymbol{\hat{\rho}},\boldsymbol{\hat{\phi}},\boldsymbol{\hat{z}} \}$.
Accordingly, the unitary transformation operator reads as $U_A={\exp[i 2 e\,(\boldsymbol{\hat{\rho}} \cdot\mathbf{A}_s)\rho/\hbar ]}=\exp[i Q_{so} x S_z ]$ and the boundary condition becomes ${\boldsymbol{\hat{\rho}}\cdot(\boldsymbol{\nabla}\hat{\mathcal{C}}')\vert_{\rho=R}=0}$.
Using this, we obtain the transformed Cooperon Hamiltonian $\hat{H}_c'$ as
\begin{align}
\hat{H}_c'/Q_{so}^2={}&\boldsymbol{\mathcal{Q}}^2-2\mathcal{Q}_z\big[\cos(Q_{so}x)S_x-\sin(Q_{so}x)S_y\big]\notag\\
&+\cos^2(Q_{so}x)S_x^2+\sin^2(Q_{so}x)S_y^2\notag\\
&-\sin(Q_{so}x)\cos(Q_{so}x)\{S_x,S_y\}+\lambda_\text{D}\mathbf{S}^2/2.
\label{HCtransformed}
\end{align}
The Dresselhaus contribution remains unchanged since $[S_i,\mathbf{S}^2]=0$.
A suitable and generic basis which satisfies the Neumann boundary condition is
\begin{align}
\braket{\mathbf{r}|n,l,Q_z}={}&J_l^{(n)}(\rho)e^{il\phi}e^{iQ_z z}/N_{nl},
\end{align}
with the angular momentum quantum number $l\in \mathbbm{Z}$, the quasi-continuous plane-wave number $Q_z$ along the wire axis, and an appropriate normalization constant $N_{nl}$.
The radial dependence is given by the Bessel function of the first kind $J_l^{(n)}$ which has its $n$-th extremum ($n\in \mathbbm{N}_+$) at the nanowire surface, i.e., $\rho=R$. 
Additionally, we  define $J_l^{(0)}=\delta_{l,0}$ which corresponds to a constant solution in the $x$-$y$ plane and constitutes the lowest mode of $H_c'$, thus, $\ket{0}\equiv \ket{n=0,l=0,Q_z}$.

\begin{figure}[t]
\includegraphics[width=.80\columnwidth]{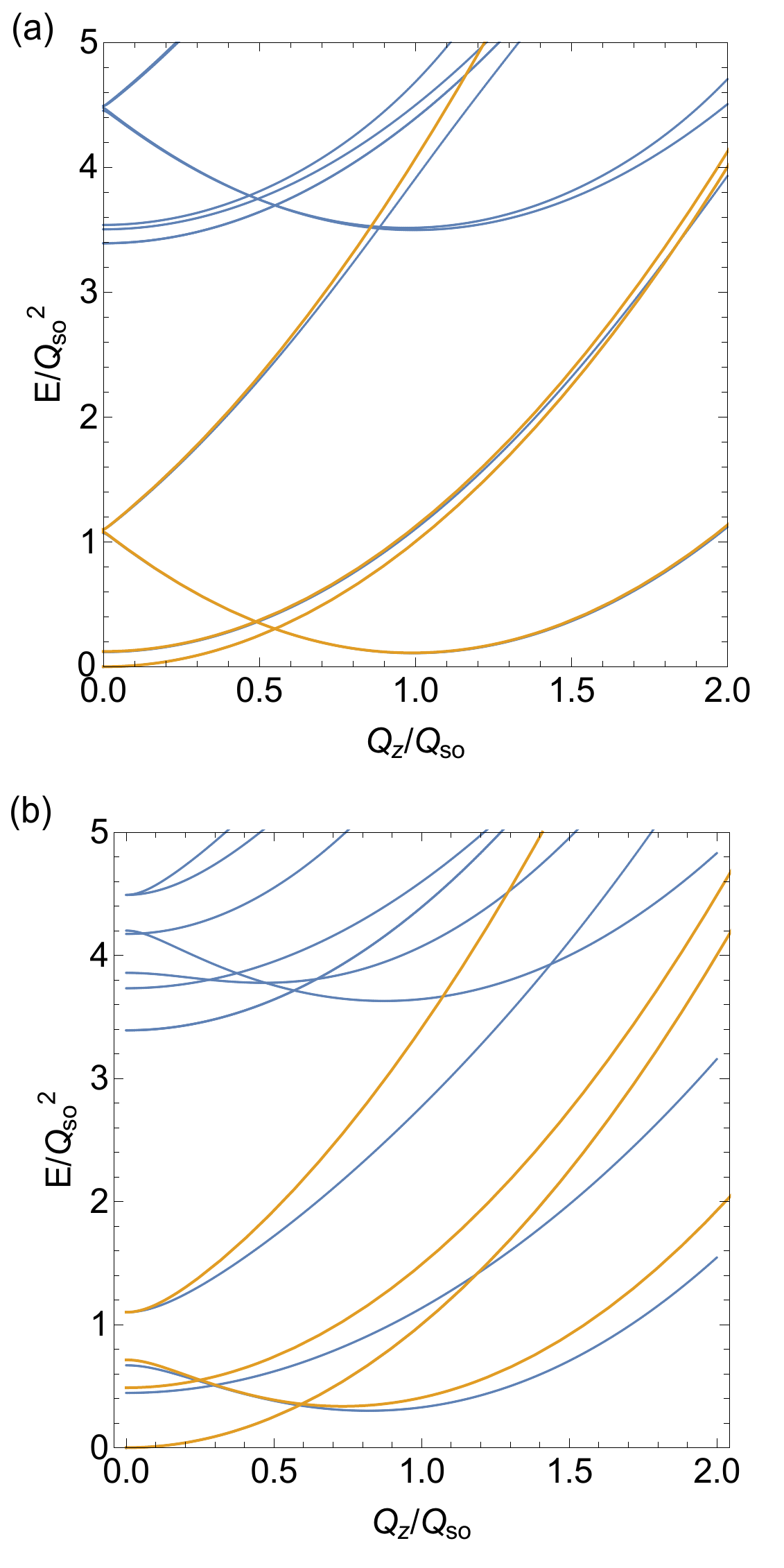}
\caption{(Color online) Comparison of the zero-mode approximation (yellow), Eqs.~(\ref{Spec1}), (\ref{Spec2}) and (\ref{Spec3}), with the exact diagonalization of $\hat{H}'_c$ (blue) truncated to $n_{max}=4$ and $|l_{max}|=4$ for $\lambda_D=0.1$ and (a) $Q_{so}R=0.3$ and (b) $Q_{so}R=1.5$.}
\label{pic:spec}
\end{figure}
%

\section{Zero-mode approximation}
In order to obtain an analytical result, the transformed Cooperon $\hat{\mathcal{C}}'$ can be evaluated only for the lowest mode $\ket{0}$.
This approach is often termed zero-mode or zero-dimensional approximation.\cite{Aleiner2001,Meyer2002,Kettemann2007a}
Using this approximation, the eigenvalues of $\braket{0|H_c'|0}$ read as
\begin{align}
E_S^{(0)}/Q_{so}^2={}&\mathcal{Q}_z^2,\label{Spec1}\\
E^{(0)}_{T,0}/Q_{so}^2={}&\mathcal{Q}_z^2+\lambda_\text{D}+a_{so}/2,\label{Spec2}\\
E^{(0)}_{T,\pm}/Q_{so}^2={}&\mathcal{Q}_z^2+\lambda_\text{D}+1-a_{so}/4\notag\\
&\pm\frac{1}{4}\sqrt{a_{so}^2+64(1-b_{so})^2\mathcal{Q}_z^2},\label{Spec3}
\end{align}
where we introduced $a_{so}=1-2J_1(2Q_{so}R)/(2Q_{so}R)$ and $b_{so}=1-2
J_1(Q_{so}R)/(Q_{so}R)$ with the Bessel function of the first kind $J_1$.
We stress that the spectrum is identical for a planar wire \cite{Kettemann2007a} if $\lambda_D=0$ and the function $2J_1(x)/x$ is replaced by $\sin (x)/x$.
In Fig.~\ref{pic:spec} we compare the zero-mode approximation with exact diagonalization for different values of $Q_{so}R$.
As the zero-mode approximation provides reliable results for small values of $Q_{so}R<1$,\cite{Wenk2010} we can write the triplet spectrum as 
\begin{align}
E^{(0)}_{T,0}/Q_{so}^2={}&\mathcal{Q}_z^2+\Delta_0,\label{SimpleSpec1}\\
E^{(0)}_{T,\pm}/Q_{so}^2={}&(\mathcal{Q}_0\pm\vert\mathcal{Q}_z\vert)^2+\Delta_1.\label{SimpleSpec2}
\end{align}
where $\Delta_0=\lambda_\text{D}+a_{so}/2$, $\Delta_1=\lambda_\text{D}+2 b_{so}-a_{so}/4$, and $\mathcal{Q}_0=1-b_{so}$.
This has the following advantages.
First, we capture the most important features of the spectrum, that is, the minima of the triplet modes $\Delta_{j}$, which are direct measures of the spin relaxation rate. 
For $E^{(0)}_{T,\pm}$ the minimum is shifted to finite momenta $\mathcal{Q}_z=\pm\mathcal{Q}_0$ and the corresponding long-lived spin densities are, therefore, of helical structure.
Second, the simple form of the spectrum allows to derive a closed-form expression for the magnetoconductance correction later on.

\section{Spin relaxation in narrow wires}\label{sec:spinrelax}
In general, four mechanisms have been found to be relevant for the relaxation of the spin of conduction electrons in metals and semiconductors: the D'yakonov-Perel', Elliott-Yafet, Bir-Aronov-Pikus, and hyperfine-interaction mechanism.\cite{Zutic2004a}
In our approach, we account only for the D'yakonov-Perel' mechanism,\cite{perel} which has shown to be a very efficient source of spin relaxation.

The D'yakonov-Perel' spin relaxation rate is related to the gaps in the Cooperon spectrum via the relation $(1/\tau_s)_j=D_e E_{T,j}$.
This is a direct consequence of the fact that there exists a unitary transformation, App.~\ref{app:relation}, which links the Cooperon with the spin diffusion equation.\cite{Wenk2010}
Hence, in general, $(1/\tau_s)_j$ depends on the Cooperon wave vector $\mathbf{Q}$  as well as on the orientation of the given spin state which is subject to a random walk.
For a spatially homogeneous spin density, $\mathbf{Q}=0$, the result is equivalent to the eigenvalues of the D'yakonov-Perel' spin relaxation tensor.\cite{Zutic2004a}

In the bulk, the 3D Cooperon can be simply evaluated in the basis of plane waves.
For $\mathbf{Q}=0$ the Cooperon Hamiltonian, Eq.~(\ref{cooperon1}), is diagonal in the basis of spin density components, App.~\ref{app:relation},  leading to the 3D spin relaxation rates $(1/\tau_s)_{ii}=D_e Q_{so}^2(\lambda_\text{D}+1+\delta_{i,y})$ where $i \in \{x,y,z\}$.
Unlike Rashba SOC, the Dresselhaus SOC affects the spin relaxation of all spin density components in the same way.

\begin{figure}[t]
\includegraphics[width=\columnwidth]{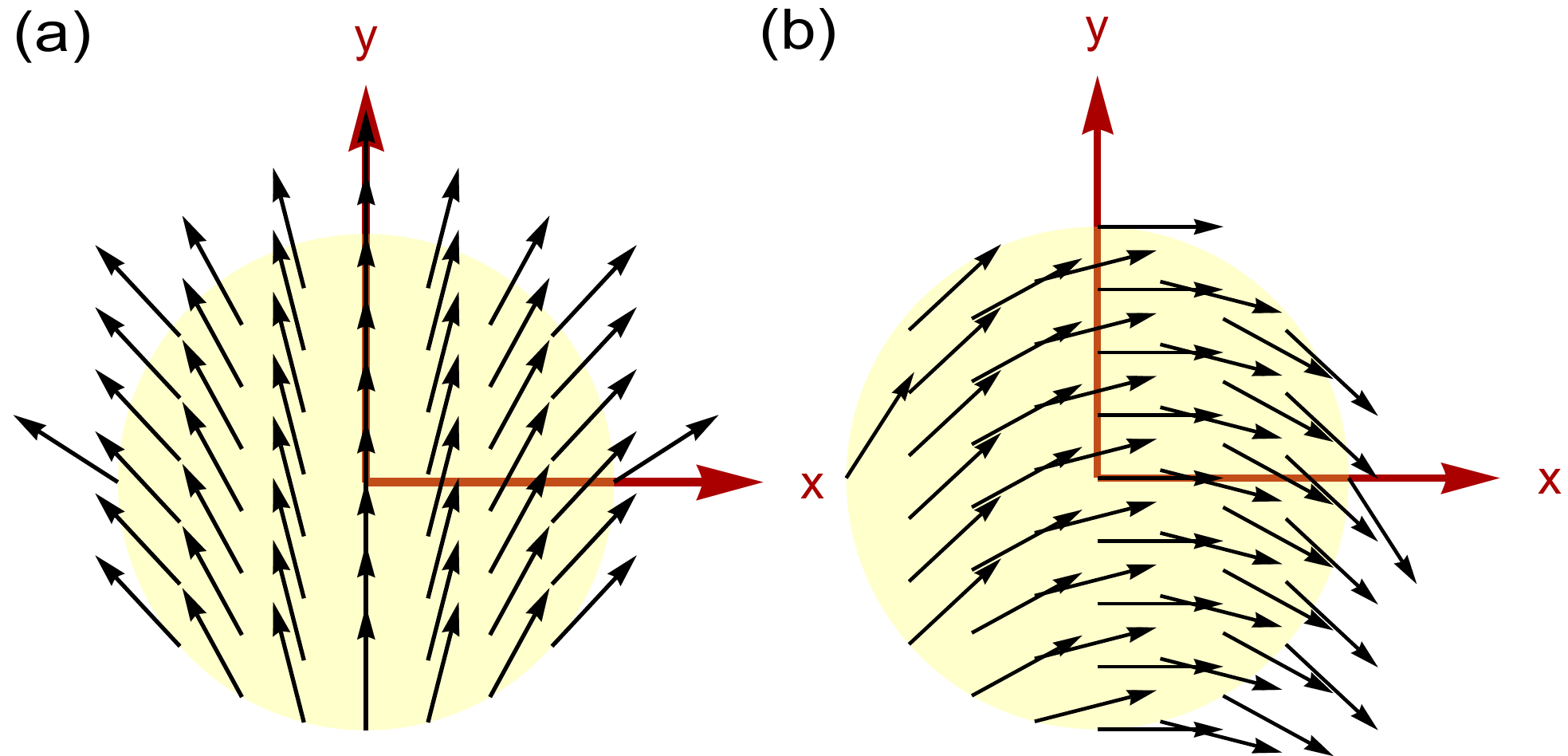}
\caption{(Color online) Eigenstates (a) $\mathbf{a}_{\parallel,-}/\Vert\mathbf{a}_{\parallel,-}\Vert$ and (b) $\mathbf{a}_{\parallel,+}/\Vert\mathbf{a}_{\parallel,+}\Vert$ of the Cooperon Hamiltonian in zero-mode approximation, App.~\ref{app:HC}, for $Q_{so}R=1$. Both states lie in the plane of the nanowire cross section (yellow).}
\label{pic:eigenstates}
\end{figure}

In presence of a radial boundary condition, the corresponding gauge transformation leads to a position dependence of the eigenstates of the Cooperon Hamiltonian $H_c$.
In App.~\ref{app:HC}, the Cooperon Hamiltonian $H_c$ in zero-mode approximation is given in the basis of spin density components.
For $Q_z=0$, the $s_z$ component is fully decoupled and independent of the location on the wire cross section.
As a consequence, a spin density which is homogeneously polarized along the wire axis is an eigenstate of the Cooperon Hamiltonian and decays according to the spin relaxation rate $(1/\tau_s)_{zz}=D_e Q_{so}^2(\lambda_\text{D}+1)$. 
Remarkably, this rate is independent of the wire radius to all orders in $Q_{so}R$ within the zero-mode approximation.
In fact, it is identical to the 3D spin relaxation rate. 
The remaining two (unnormalized) eigenstates  $\mathbf{a}_{\parallel,j}$, where $\mathbf{a}_{\parallel,-}=(\tan(Q_{so}x),1,0)^\top$ and $\mathbf{a}_{\parallel,+}=(1,-\tan(Q_{so}x),0)^\top$, with the according spin relaxation rates $(1/\tau_s)_{\parallel,j}=D_e Q_{so}^2(\delta_{j,-}\pm a_{so}/2+\lambda_\text{D})$ lie in the plane of the cross section and depend on the position as depicted in Fig.~\ref{pic:eigenstates}.

In analogy to numerous previous works\cite{Dyakonov1986,Iordanskii1994,Pikus1995,Knap1996,Kettemann2007a} we define hereafter the spin relaxation rate of the system $1/\tau_s$ as the minimal rate for $\mathcal{Q}_z=0$, which is $1/\tau_s\equiv(1/\tau_s)_{\parallel,+}=D_eQ_{so}^2(a_{so}/2+\lambda_\text{D})$.
In the limit of $Q_{so}R\ll 1$, corresponding to a radius $R$ much smaller than the Rashba spin precession length $L_{so}$, we replace $a_{so}\rightarrow 4b_{so}\rightarrow (Q_{so}R)^2/2$, which gives,
\begin{align}
\frac{1}{\tau_s}={}&\frac{k_F^2\alpha_\text{R}^2\tau_e}{3\hbar^2}(Q_{so}R)^2+\frac{32 k_F^6\gamma_D^2\tau_e}{105\hbar^2}
\label{SRrate}
\end{align}
to second order in $Q_{so}R$. 
Noting further that since $Q_{so}x \leq Q_{so}R\ll 1$, the respective eigenstate is $\mathbf{a}_{\parallel,+}\approx(1,0,0)^\top$.
In accordance with Ref.~\onlinecite{Kettemann2007a}, the first term in Eq.~(\ref{SRrate})  is strongly suppressed in wires with small radii.
However, compared to Ref.~\onlinecite{Kettemann2007a} the first term is a factor 2 smaller if we associated $R=W/2$, where $W$ is the width of the planar quantum wire. 
The Dresselhaus-dependent spin relaxation rate was also obtained in Ref.~\onlinecite{PikusTitkov2}.
Notably, as seen from Eq.~(\ref{SimpleSpec2}), the global minimum of the spectrum is found at finite wave vectors $\mathcal{Q}_z=\pm\mathcal{Q}_0$ and given by $\Delta_1$ which is for small $\lambda_\text{D}$ approximately half as large as $\Delta_0$.
This outlines the superior spin-lifetime of helical spin-densities, which was observed earlier in planar and tubular  two-dimensional electron gases (2DEGs).\cite{Kettemann2007a,Kammermeier2016,Kammermeier2016PRL}

\section{General remarks}

First, we would like to point out some general observations on the structure of the Cooperon Hamiltonian $\hat{H}_c$, Eq.~(\ref{3DCooperonHamiltonian}), with respect to the presence of a generic SOC contribution.
Linear and cubic SOC terms can be always expanded in terms of first- and third-degree spherical harmonics in the wave vector $\mathbf{k}$.
Only the first-degree spherical harmonic terms can be rewritten in form of an effective vector potential $\mathbf{A}_s$ in the Cooperon Hamiltonian.
In the present case, e.g., the bulk Dresselhaus SOC consists only of third-degree  spherical harmonic terms, which leads to the structure of Eq.~(\ref{3DCooperonHamiltonian}). 
Owing to the effective vector potential, the minimum of $\hat{H}_c$ is shifted to finite Cooperon wave vectors $\mathbf{Q}_\text{min}$.

This has important consequences when the Cooperon Hamiltonian is subject to an insulating spin-conserving boundary condition as a result of a finite-size geometry (cf. Sec.~\ref{sec:boundary}).
In particular, the component of the effective vector field  $\mathbf{A}_s$ which is perpendicular to the surface is removed by a gauge transformation $U_A$.
Thereby, the component of the minimum in direction of the boundary, i.e., $\mathbf{\hat{n}}\cdot\mathbf{Q}_\text{min}$, is shifted to zero wave vector after the transformation, i.e., $U_A(\mathbf{\hat{n}}\cdot\mathbf{Q}_\text{min})U_A^\dag=0$.
Since the lowest Cooperon mode typically corresponds to a constant solution in coordinate space, $\braket{0|\mathbf{\hat{n}}\cdot\mathbf{Q}|0}=0$ holds true and the remaining gauge-transformed (and therefore position-dependent) terms are averaged along the confined directions.
This generates a suppression of the spin relaxation rate in small wires (in zero-mode approximation), which is denoted as motional narrowing as it is observed in the cylindrical wire in this work and also earlier in planar quantum wires.\cite{Kettemann2007a,Wenk2010,Wenk2011}
In both cases, the spin relaxation due to first-degree  spherical harmonic SOC terms is strongly suppressed for wires of widths much smaller than the spin precession length.
A finite relaxation remains, however, due to the third-degree  spherical harmonic contributions.

Hence, the effect of motional narrowing is expected to be maximal if the entire effective vector potential $\mathbf{A}_s$ is removed.
Here, the cylindrical nanowire can take an outstanding role as it has a  boundary in radial direction for the Cooperon, in contrast to the planar quantum wire that has a boundary along one Cartesian coordinate only.

\section{Magnetic dephasing}\label{sec:mag}
For experimental probing, we need to take into account the phase breaking 
due to a magnetic field. 
Disregarding SOC, the Cooperon propagator $\hat{\mathcal{C}}$ in real space is defined through the diffusion equation \cite{AltshulerAronov1981,Meyer2002}
\begin{align}
\left[\hbar D_e(i\boldsymbol{\nabla}-2e\mathbf{A}/\hbar)^2+\hbar/\tau_\phi\right]\hat{\mathcal{C}}(\mathbf{r}-\mathbf{r'})={}&\delta(\mathbf{r}-\mathbf{r'}),
\end{align}
with the magnetic vector potential $\mathbf{A}$ and the dephasing rate $\tau_\phi$.
This equation has the general solution
\begin{align}
\hat{\mathcal{C}}(\mathbf{r}-\mathbf{r'})={}&\sum_n\Phi_n^*(\mathbf{r}')\Phi_n(\mathbf{r})/(\hbar D_e E_n+\hbar/\tau_\phi),
\end{align}
where $\Phi_n$ solve the eigenvalue equation $(i\boldsymbol{\nabla}-2e\mathbf{A}/\hbar)^2\Phi_n(\mathbf{r})=E_n\Phi_n(\mathbf{r})$ with the according eigenenergy $E_n$.
If we choose a gauge such that the vector potential has no component perpendicular to the surface, i.e., $\mathbf{\hat{n}}\cdot\mathbf{A}=0$, the vector potential does not affect the  boundary condition, Eq.~(\ref{boundary}).
For small magnetic fields, we can treat the terms $\propto\mathbf{A}$ in the eigenvalue equation perturbatively in zero-mode approximation. 
Assuming a Coulomb gauge with $\braket{0|\mathbf{A}|0}=0$,
we obtain in lowest order the magnetic phase shift rate as
\begin{align}
1/\tau_B={}&D_e\left(2e/\hbar\right)^2\braket{0|\mathbf{A}^2|0},
\end{align}
where the expectation value is equivalent to the average taken over the sample geometry.
The same expression is found by Beenakker \textit{et al.} in Ref.~\onlinecite{Beenakker1988a}.

If a cylindrical nanowire with the surface vector $\mathbf{\hat{n}}=\boldsymbol{\hat{\rho}}$ is placed in a magnetic field perpendicular ($\mathbf{B}_\perp=B\mathbf{\hat{y}}$) or parallel ($\mathbf{B}_\parallel=B\mathbf{\hat{z}}$) to  the growth axis $\mathbf{\hat{z}}$, the corresponding vector potentials that fulfill the above-mentioned criteria are $\mathbf{A}_\perp=By \mathbf{\hat{z}}$ and $\mathbf{A}_\parallel=B(x \mathbf{\hat{y}}-y \mathbf{\hat{x}})/2$.
Consequently, the respective magnetic phase shift rates become $1/\tau_{B,\perp}=D_e\left(e B R/\hbar\right)^2$ and $1/\tau_{B,\parallel}=1/\left(2\tau_{B,\perp}\right)$.
Note that here the magnetic field is assumed to be small enough such that the magnetic length\cite{AltshulerAronov1981} $l_B=\sqrt{\hbar/(2e\vert B\vert)}$ exceeds the wire width.

\section{Magnetoconductance correction}
For cylindrical semiconductor nanowires of length $L$ and radius $R$, the macroscopic magnetoconductance correction $\Delta G$ follows the relation $\Delta G=(\pi R^2/L) \Delta  \sigma$.
More explicitly, using the groundwork of the preceding paragraphs we find in zero-mode approximation
\begin{align}
\Delta \sigma^{(0)}(B)={}&\frac{2e^2}{h}\frac{1}{R^2 \pi^2Q_{so}}\int_0^{\sqrt{c_e}}d\mathcal{Q}_z\Bigg(\frac{1}{E_S^{(0)}/Q_{so}^2+c_\phi+c_B}\notag\\
&-\sum_{j\in\{0,\pm\}}\frac{1}{E_{T,j}^{(0)}/Q_{so}^2+c_\phi+c_B}\Bigg),
\label{conductivity}
\end{align}
with $c_i=1/(D_e Q_{so}^2\tau_i)$ where $i\in\{e,\phi,B\}$.
The exact result for the magnetoconductance correction is obtained by analogously summing over all eigenmodes of the Cooperon Hamiltonian, that is, $\Delta G=\sum_n \Delta G^{(n)}$.
We stress that the upper cutoff $\sqrt{c_e}$ due to the scattering rate $\tau_e$, to remove the divergence, is strictly speaking only required in 2D.
Thus, neglecting the upper limit $\sqrt{c_e}$ and using the simplified triplet spectrum for $Q_{so}R<1$, Eqs.~(\ref{SimpleSpec1}) and (\ref{SimpleSpec2}), we obtain  the closed-form expression
\begin{align}
\Delta \sigma^{(0)}(B)={}&\frac{2e^2}{h}\frac{1}{2\pi Q_{so} R^2 }\Bigg(\frac{1}{\sqrt{c_\phi+c_B}}\notag\\
&-\frac{2}{\sqrt{\Delta_1+c_\phi+c_B}}-\frac{1}{\sqrt{\Delta_0+c_\phi+c_B}}\Bigg),
\end{align}
which resembles the result in Ref.~\onlinecite{Kettemann2007a}.
In the following, we apply the developed model to fit  magnetoconductance measurements.

%
\section{Experimental data fitting}\label{sec:exp}
Exemplarily, we present the fitting results for a heavily \textit{n}-doped InAs nanowire.
As previously shown,\cite{Heedt2015,Kammermeier2016,Degtyarev2017} the electrons in undoped InAs nanowires are confined to a narrow layer beneath the surface due to Fermi level pinning and the transport is governed by surface states.
However, a controlled doping allows the electrons to distribute over the entire volume and thereby change the dimensionality and transport topology to that of a quasi-3D channel. \cite{Wirth2011,Heedt2015,Degtyarev2017}

The studied sample corresponds to \textit{Device D} of Ref.~\onlinecite{Heedt2015} and possesses the following parameters.
Adopting the findings of Ref.~\onlinecite{winklerbook}, the narrow band gap of InAs results in large Rashba and Dresselhaus SOC coefficients $\gamma_R=\SI{117.1}{e\angstrom^2}$ and $\gamma_D=\SI{27.18}{eV\angstrom^3}$, respectively.
Moreover, the effective mass is given by $m=0.026 \,m_\text{e}$ where $m_\text{e}$ is the bare electron mass. \cite{Bringer2011}
In line with the experimental setup of Ref.~\onlinecite{Heedt2015}, we consider a length of the nanowire of $L=\SI{2.18}{\micro m}$ and a radius of $R=\SI{47.5}{nm}$.
Moreover, we use the field-effect mobility $\mu=\SI{600}{cm^2V^{-1}s^{-1}}$ and the 3D electron density $n_{3D}=\SI{5e18}{cm^{-3}}$.
The change of the back-gate voltage $V_g$ from $\SI{5}{V}$ to $\SI{60}{V}$ yields an increase of the electron density by a factor of 1.5, whereas the mobility is assumed to remain relatively unchanged.
By means of the relations $\mu=e\tau_e/m$ and $k_F=(3\pi^2n_{3D})^{1/3}$, we find a mean free path $l_{e}=v_F\tau_e$ between $\SI{21}{nm}$ and $\SI{24}{nm}$.
Accordingly, $\hbar/(\epsilon_F\tau_e)$ ranges from 0.14 to 0.18 and therefore the \textit{Ioffe-Regel} criterion is generally well fulfilled.

\begin{figure}[t]
\includegraphics[width=\columnwidth]{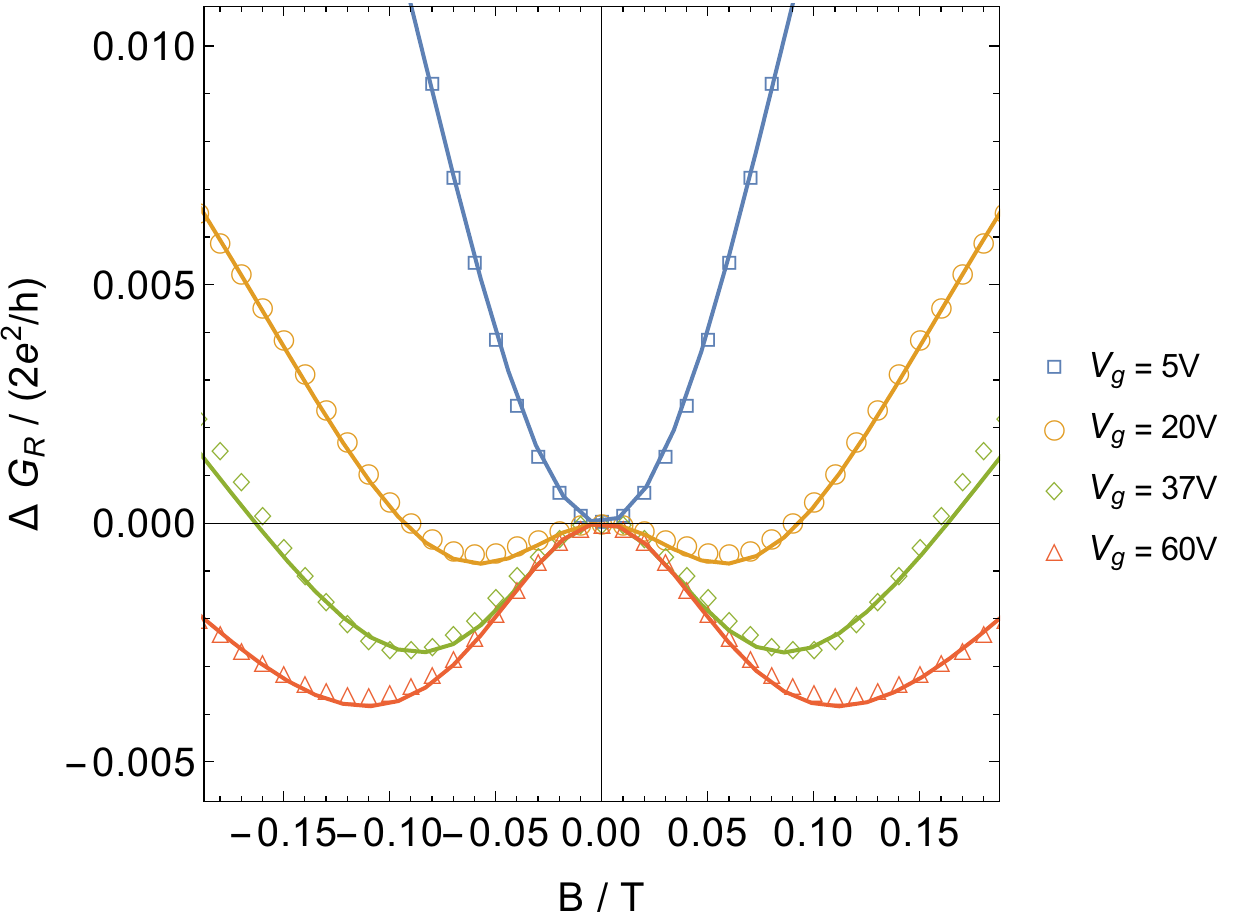}
\caption{(Color online) Gate-controlled crossover from positive to negative relative magnetoconductance $\Delta G_R\equiv\Delta G(B)-\Delta G(B=0)$ in a doped $\braket{111}$ InAs nanowire. The symbol-dotted lines correspond to experimental data for different back-gate voltages $V_g$, which are fitted by theory (solid lines) using Eq.~(\ref{conductivity}) and varying the Rashba and effective Dresselhaus SOC strengths as shown in Fig.~\ref{pic:FitPlotPaper2}(a).}
\label{pic:FitPlotPaper1}
\end{figure}

\begin{figure}[t]
\includegraphics[width=\columnwidth]{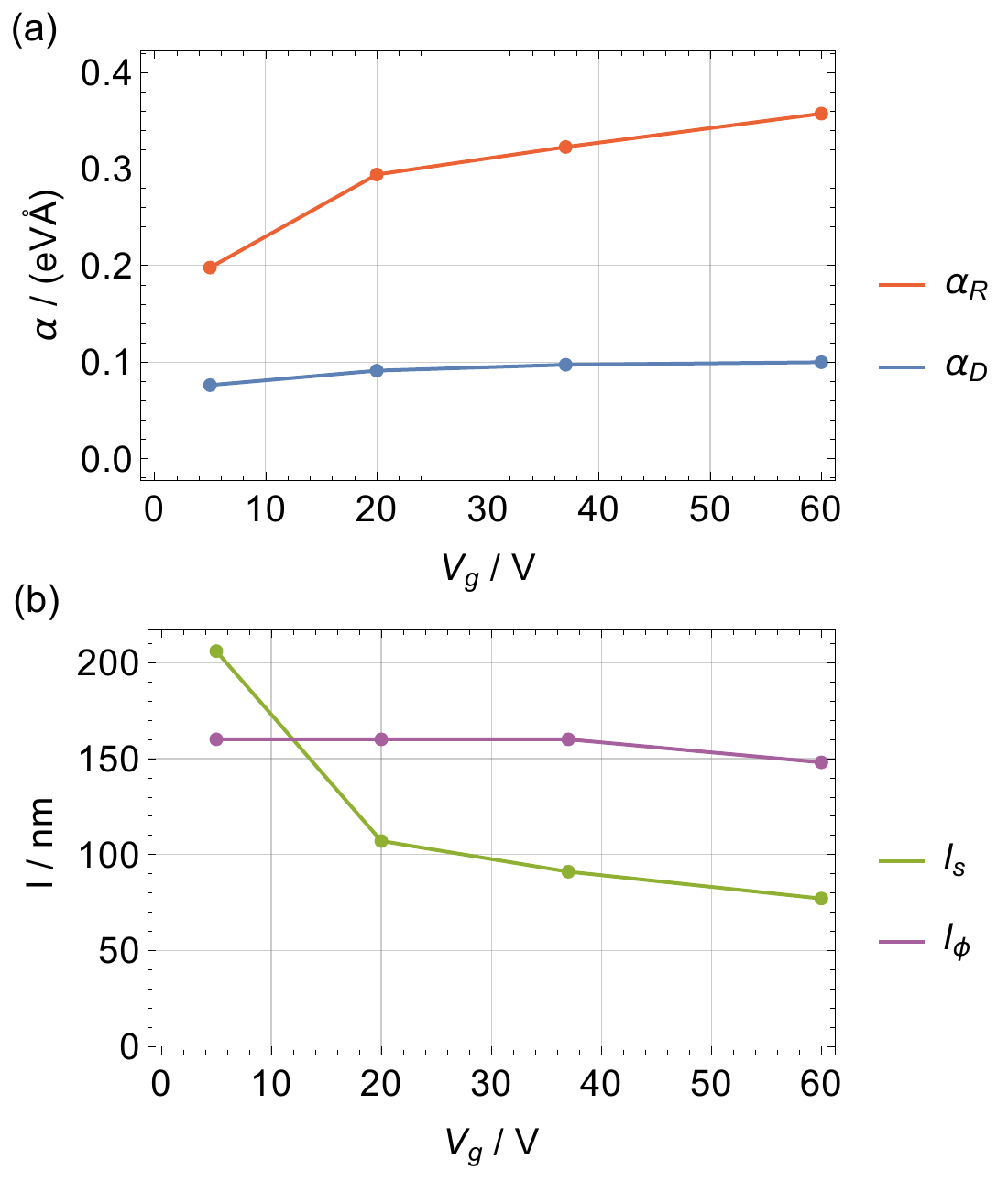}
\caption{(Color online) Extracted fitting parameters for a doped $\braket{111}$ InAs nanowire. We show in (a) the Rashba and effective Dresselhaus SOC strength $\alpha_R$ and $\alpha_D=\gamma_Dk_F^2$, and in (b) the spin relaxation and dephasing length $l_s$ and $l_\phi$, respectively, in dependence of the back-gate voltage $V_g$.}
\label{pic:FitPlotPaper2}
\end{figure}

Fig.~\ref{pic:FitPlotPaper1} depicts a gate-induced crossover from positive to negative relative magnetoconductance $\Delta G_R\equiv\Delta G(B)-\Delta G(B=0)$, which is usually associated with a crossover from weak localization to weak antilocalization. 
The experiments are performed at a temperature of $T=\SI{4}{K}$ and the magnetic field is oriented perpendicular to the wire axis, i.e., $\tau_B=\tau_{B,\perp}$ (cf. Sec.~\ref{sec:mag}).
In order to average out the superimposed universal conductance  oscillations, each  magnetoconductance curve represents the mean value of roughly 200 individual measurements in $\SI{20}{V}$ gate voltage intervals.  
We fitted Eq.~(\ref{conductivity}) by changing the effective Dresselhaus parameter $\alpha_D=\gamma_D k_F^2$ according to the modifications of the electron density and by adjusting the Rashba parameter $\alpha_R$, or equivalently the strength of the internal electric field $\vert\mathcal{E}\vert$.
The resulting electric field increases with the gate voltage from $\SI{1.7e7}{V/m}$ to $\SI{3.1e7}{V/m}$.
The Rashba and effective Dresselhaus SOC strengths are shown in Fig.~\ref{pic:FitPlotPaper2}(a).
A slight deviation from the typically expected linear $V_g$-dependence of $\alpha_R$ is attributed to deviations from a homogeneous electric field within the wire.
We point out, that the precondition for the zero-mode approximation, $Q_{so} R<1$, is strictly speaking not perfectly fulfilled for large voltages.
More precisely, $Q_{so} R$ ranges from $0.64$ to $1.16$.
However, by comparing the exact diagonalization with the zero-mode approximation, cf. Fig.~\ref{pic:spec}, it becomes obvious that the most important characteristics of the spectrum, the minima, are barely changed and the application of the zero-mode approximation is here still justified.
Furthermore, using the relation $l_i=\sqrt{D_e\tau_i}$, $i\in\{s,\phi\}$ and $1/\tau_s\equiv(1/\tau_s)_{\parallel,+}$ as defined in Sec.~\ref{sec:spinrelax} (and in the limit $Q_{so}R\ll 1$ in Eq.~(\ref{SRrate})), we can extract the spin relaxation and dephasing lengths, $l_s$ and $l_\phi$, respectively, which are shown in Fig.~\ref{pic:FitPlotPaper2}(b).
At $V_g=\SI{5}{V}$ the spin relaxation length $l_s$ exceeds the dephasing length $l_\phi$, which reflects the observation of weak localization in Fig.~\ref{pic:FitPlotPaper1}.
Hence, a controlled application of a gate voltage allows to reduce the spin relaxation length roughly by a factor of 3.

\subsection{Critical Discussion and Comparison with Previous Results}

Hereafter, we follow with a critical discussion and compare our gathered data with previous experiments on similar \textit{n}-doped InAs nanowire devices.\cite{Roulleau2010,Liang2012,Scheruebl2016,Takase2017}
In these works, the magnetoconductance data are analyzed by means of a 1D magnetoconductance formula,\cite{Kurdak1992} which does not consider the mesoscopic details of the system.
In App.~\ref{app:kurdak}, we use this formula to fit the magnetoconductance data of our device.
Aside from the obvious discrepancy between experimental data and theory, it shows disagreements with our findings above. 

First, we remark that even for a vanishing back-gate voltage a finite Rashba strength $\alpha_R$ will remain,
which was also seen in Refs.~\onlinecite{Roulleau2010,Liang2012,Scheruebl2016,Takase2017}.
We attribute this to the fact that even for zero gate voltage an intrinsic electric field due to Fermi level surface pinning will remain.
Aside from that, for small voltages other spin relaxation mechanisms can become important, above all, the Elliott-Yafet (EY) mechanism due to the large electron density through doping.
However, we emphasize that the modification of the electron density in our sample alters the EY spin relaxation rate by a factor of 1.7. \cite{Chazalviel1975}
Since we detect an increase of the spin relaxation rate by a factor of 9.5, it is not possible to explain this behavior within EY theory. 
Nevertheless, the strength of the extracted Rashba parameter $\alpha_R$ should be treated with caution as it comprises a contribution of additional spin relaxation rates.


Secondly, although similar transport parameters are found in Refs.~\onlinecite{Roulleau2010,Liang2012,Scheruebl2016,Takase2017}, the variation of the dephasing length $l_\phi$ with the gate voltage in the according fits is most striking.
In most of these works, the dephasing length increases with the gate voltage which is usually justified by a reduced electron-electron interaction through an increase of the electron density.
Some authors even observed a decrease of $l_\phi$ with increasing gate voltage\cite{Scheruebl2016} or oscillations.\cite{Roulleau2010}
It is most pronounced in Refs.~\onlinecite{Roulleau2010,Liang2012,Takase2017}, where the dephasing length suddenly changes by about $\SI{100}{nm}$ near the WL regime.
In App.~\ref{app:kurdak}, we show that the application of the 1D magnetoconductance formula of Ref.~\onlinecite{Kurdak1992} to our nanowire device likewise leads to an unusual trend for $l_\phi$.
This behavior is not seen in our device with our magnetoconductance model, where the dephasing length remains nearly constant.
However, we find that in our model an unambiguous fitting of the magnetoconductance curve in the WL regime, in contrast to the WAL regime and opposed to the model in Ref.~\onlinecite{Kurdak1992}, is barely possible.
Note that we could also fit for a lower value of $l_\phi$ for $V_g=\SI{5}{V}$ which would further increase the spin relaxation length and diminish the saturation value for $\alpha_R$.
However, as we do not see any indication of a change of $l_\phi$ in the WAL regime, we assume that a similar value holds in the WL regime.
This finding supports the need of taking into account details on the mesoscopic scale of the nanowire as presented in this paper in order to obtain reliable transport parameters. 


%
%
%
%
%
%
%

Based on our observations, we suggest that for dephasing the electron-electron interaction may be not as effective as previously assumed in a largely doped sample as considered here.
This would be in agreement with the findings in disordered 3D metal films.\cite{Lin2002}
On the other hand, the change of electron density in our investigated system is possibly too low in order to make a reliable statement.
Also, as the extracted dephasing length exceeds the diameter $d$ of the wire, we expect the geometric properties to play an important role
in a similar manner as it is the case for the magnetic dephasing.
In planar quantum wires with a width smaller than the dephasing length electron-electron interaction has been identified as the predominant mechanism.\cite{Choi1987}
For further studies, we propose therefore (a) the development of a theoretical description of the inelastic scattering mechanisms as a function of temperature, electron density, and system size for a quasi-3D cylindrical wire and (b) an experimental investigation to see which mechanisms really apply.
This would support a reliable  parameter fitting in the WL regime and thereby enable a correct determination of the zero gate voltage spin relaxation processes.


\section{Summary and perspective}

We studied the effects of SOC on the quantum conductivity correction for semiconductor nanowires with zinc-blende structure.
The spin relaxation due to Dresselhaus SOC is found to be the same for all spin components, independent of the wire growth direction and the wave vector of the spin density, and not affected by a change of the wire radius.
Contrarily, in presence of Rashba SOC the relaxation depends on the spin component.
A homogeneous spin density that is polarized along the $\mathbf{\hat{x}}$ axis decays according to Eq.~(\ref{SRrate}) if the wire radius is smaller than the spin precession length.
However, the long-lived spin states have helical structure in real space.
Similarly to the planar wire,\cite{Kettemann2007a} the relaxation due to Rashba SOC is strongly suppressed for small wire widths.
Interestingly, a homogeneously excited spin density along the wire axis does not exhibit any dependence on the wire radius and is therefore not subject to motional narrowing.
 The derived expressions for the magnetoconductance correction are fitted to the data of magnetoconductance measurements of a heavily doped back-gated InAs nanowire.
We find good agreement between theory and experiment and reasonable transport parameters.
For comparison, we also apply the 1D magnetoconductance formula of Kurdak \textit{et al.},\cite{Kurdak1992} which has been frequently used by other authors.\cite{Hansen2005,Dhara2009,XiaoJie2010,
Weperen2015,Roulleau2010,Liang2012,Scheruebl2016,Takase2017}
The fitted curves show larger deviations from the experimental observations and an unusual trend of the dephasing length.

We stress that the developed model holds for 3D-diffusive nanowires and a crossover to the quasi-ballistic regime is not included.
For the latter case, it is plausible to assume that the Dresselhaus spin relaxation rate will decrease due to the reduction of the number of contributing channels as shown in Ref.~\onlinecite{Wenk2011} for planar wires.
Additionally, the effects of surface roughness will start to play a noticeable role.\cite{Khalaf2016}
It is also to mention that semiconductor nanowires are often polytypic with zinc-blende and wurtzite segments or even pure wurtzite phase, even though the underlying semiconductor material has zinc-blende lattice in the bulk. \cite{Hiruma1995,ZhangLu2014}
As the SOC in the wurtzite phase is fundamentally different, distinct characteristics concerning conductivity and spin relaxation can be expected and therefore further model calculations are strongly requested.

\section{Acknowledgements}
We gratefully acknowledge K. Sladek and H. Hardtdegen for nanowire growth and we thank I. Otto for valuable experimental support.
The authors also thank F. Dirnberger, D. Bougeard, A. Bringer, F. Mireles, and S. Csonka for useful discussions.
This work was supported by Deutsche Forschungsgemeinschaft via Grants No. SFB 689 and No. FOR 912.

\appendix
\section{Relation  between triplet basis and spin density components}\label{app:relation}

As shown in Ref.~\onlinecite{Wenk2010}, there exists a unitary transformation between the spin diffusion equation and the Cooperon.
Therefore, we obtain an according transformation between the spin density $\mathbf{s}=(s_x,s_y,s_z)^\top$ and the triplet vector $\mathbf{\tilde{s}}=(\ket{1,1},\ket{1,0},\ket{1,-1})^\top$  of the Cooperon, which is
\begin{align}
\mathbf{\tilde{s}}&={}U_{cd}\,\mathbf{s},
\end{align}
with the unitary operator
\begin{align}
U_{cd}&={}\begin{pmatrix}
-1&i&0\\
0&0&\sqrt{2}\\
1&i&0
\end{pmatrix}/\sqrt{2}.
\end{align}

\section{Spin matrices}\label{app:spin}

The spin matrices of a system with two electrons in singlet-triplet basis $\ket{s,m_s}$, with total spin quantum number $s \in\{0,1\}$ and according magnetic quantum number $m_s \in\{0,\pm1\}$, are

\begin{align}
S_x={}&
\,\frac{1}{\sqrt{2}}
\begin{pmatrix}
0&0&0&0\\
0&0&1&0\\
0&1&0&1\\
0&0&1&0
\end{pmatrix},\notag\\ 
S_y={}&
\,\frac{i}{\sqrt{2}}
\begin{pmatrix}
0&0&0&0\\
0&0&-1&0\\
0&1&0&-1\\
0&0&1&0
\end{pmatrix},\notag\\ 
S_z={}&
\,
\begin{pmatrix}
0&0&0&0\\
0&1&0&0\\
0&0&0&0\\
0&0&0&-1
\end{pmatrix},
\label{spinst}
\end{align}
in the order $\{\ket{0,0},\ket{1,1},\ket{1,0},\ket{1,-1}\}$. 
The singlet and triplet sectors are decoupled in this representation.

\section{Cooperon Hamiltonian in zero-mode approximation in terms of spin density components}\label{app:HC}

Relevant in experiments is the relaxation process of the spin density $\mathbf{s}$.
Due to the gauge transformation $U_A$ the eigenstates of the Cooperon Hamiltonian $H_c$ depend on the position on the cross section.
We can write the triplet sector of the Cooperon Hamiltonian in the basis of the spin density components by reverting the gauge transformation after projecting the transformed Cooperon Hamiltonian on the zero-mode and applying the basis transformation to the triplet sector as defined in App.~\ref{app:relation}.
More precisely, the triplet sector of the Cooperon Hamiltonian  in the basis of the spin density components $\{s_x,s_y,s_z\}$ reads in terms of $Q_{so}^2$ as
\begin{align}
\begin{pmatrix}
a& d&e\\
d^*&b&f\\
e^*&f^*&c
\end{pmatrix},
\end{align}
where
\begin{align}
a={}&c-\frac{1}{2}+\frac{1}{2}(a_{so}-1)\cos(2Q_{so}x),\notag\\
b={}&c-\frac{1}{2}-\frac{1}{2}(a_{so}-1)\cos(2Q_{so}x),\notag\\
c={}&\mathcal{Q}_z^2+\lambda_\text{D}+1,\notag\\
d={}&-\frac{1}{2}(a_{so}-1)\sin(2Q_{so}x),\notag\\
e={}&-2i \mathcal{Q}_z(b_{so}-1)\sin(2Q_{so}x),\notag\\
f={}&-2i \mathcal{Q}_z(b_{so}-1)\cos(2Q_{so}x).
\end{align}
Notably, for $\mathcal{Q}_z=0$ the $s_z$ component is decoupled and independent of the location on the wire cross section and the wire radius.
Consequently, a spin density which is homogeneously polarized along the wire axis is not subject to motional narrowing in zero-mode approximation. 

\section{Experimental data fitting with Kurdak et al.'s formula}\label{app:kurdak}

\begin{figure}[b]
\includegraphics[width=\columnwidth]{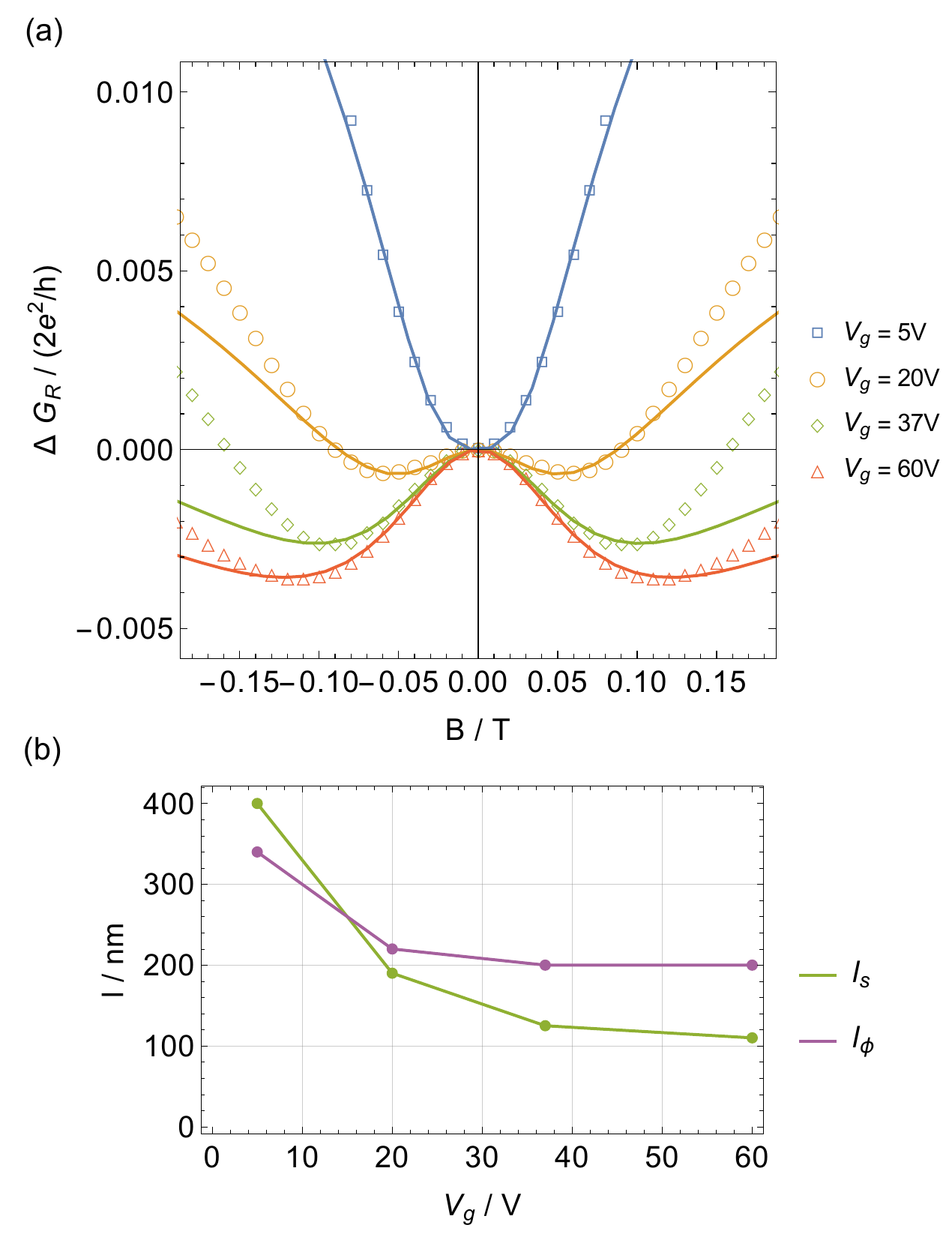}
\caption{(Color online) (a) Gate-controlled crossover from positive to negative relative magnetoconductance $\Delta G_R\equiv\Delta G(B)-\Delta G(B=0)$ in a doped $\braket{111}$ InAs nanowire. The symbol-dotted lines correspond to experimental data for different back-gate voltages $V_g$, which are fitted by the 1D magnetoconductance formula of Kurdak \textit{et al.}\cite{Kurdak1992} (solid lines), and adjusting the dephasing and spin relaxation lengths, $l_\phi$ and $l_s$, as shown in (b).}
\label{pic:Kurdak}
\end{figure}

Here, we demonstrate the application of the 1D magnetoconductance formula of Kurdak \textit{et al.}, Ref.~\onlinecite{Kurdak1992}, to the nanowire device discussed in Sec.~\ref{sec:exp}.
This model is frequently used for the theoretical analysis of semiconductor nanowire devices.\cite{Hansen2005,Dhara2009,XiaoJie2010,
Weperen2015,Roulleau2010,Liang2012,Scheruebl2016,Takase2017}
In case of a diffusive wire  of length $L$, the magnetoconductance correction reads as
\begin{align}
\Delta G(B)={}&\frac{2e^2}{h}\frac{1}{2 L }\Bigg[3\left(\frac{1}{l_\phi^2}+\frac{4}{3l_s^2}+\frac{1}{l_B^2}\right)^{-1/2}\notag\\
&-\left(\frac{1}{l_\phi^2}+\frac{1}{l_B^2}\right)^{-1/2}\Bigg],
\end{align}
with the dephasing, spin relaxation, magnetic dephasing length, $l_\phi$, $l_s$, and $l_B$, respectively.
For the magnetic dephasing length, we used our relation for a perpendicular magnetic field, that is, $l_B=\sqrt{D_e\tau_{B,\perp}}$  as derived in Sec.~\ref{sec:mag} which is more appropriate for a cylindrical wire.
Note that compared to the definition in Refs.~\onlinecite{Chakravarty1986,Beenakker1988a}, here also $l_B\propto \vert B\vert^{-1}$ holds true.
In Fig.~\ref{pic:Kurdak} we show the relative magnetoconductance correction $G_R=\Delta G(B)-\Delta G(0)$ and the accordingly obtained fitting parameters.
Remarkably, in strong contradiction to the observations using our model, cf. Sec.~\ref{sec:exp}, the extracted dephasing length shows a monotonous decrease with the gate voltage which is rather unphysical.
Also, the spin relaxation length is nearly twice as large for small gate voltages.
Aside from that, a strong discrepancy between the experimental data and theory in the weak antilocalization regime is obvious.
As a consequence of these observations, we suggest that 
a more appropriate model should be used.

\bibliographystyle{apsrev4-1}
\bibliography{WK}
\end{document}